\documentclass[10pt,twocolumn]{article}
\usepackage[margin=1in]{geometry}
\usepackage{amsmath}
\usepackage{abstract}
\usepackage{amsfonts,color}
\usepackage{amssymb}
\usepackage{graphicx}
\usepackage{hyperref}
\usepackage{mathrsfs}
\usepackage{bigints}
\begin{document}
\title{\bf{Instabilities in nonrelativistic spherically symmetric self-gravitating accretion}}
\author
{Satadal Datta\\
Harish-Chandra research Institute, Chhatnag Road, Jhunsi, Prayagraj-211019, HBNI, INDIA\\
Email: satadaldatta1@gmail.com, satadaldatta@hri.res.in}
\twocolumn[
\maketitle
\begin{onecolabstract}
We consider time dependent problem in perturbative approach for a nonrelativistic inviscid spherically symmetric accretion model where the effect of the gravity of the medium is considered in Newtonian gravity framework. We consider spherically symmetric nonrelativistic accretion, i.e. Bondi accretion with self-gravity. Our approach is perturbative in the linear order of perturbation regime. We introduce linear perturbation over the existing steady state solution of the system. The analysis has two features, one is that the linear perturbation in mass accretion rate in such irrotational inviscid model of accretion gives rise to emergent gravity and on the other hand, we get some significant insights about instabilities in the flow due to the effect of gravity in the medium, whereas the instabilities are absent in the absence of self-gravity.  
\end{onecolabstract}
]
\vspace*{0.33cm}
\section{Introduction}
There are existing works \cite{b}\cite{c} on the correction over the steady state solutions of Bondi accretion \cite{a} due to the effect of gravity of the accreting medium. We linearize density, velocity of the flow to analyse the stability of such steady state solutions. Transonic accretion for inviscid irrotational accretion models are natural systems where analogous blackhole horizon like effect, i.e. acoustic 'dumbhole' appears to be an emergent phenomena through linear perturbations in certain quantities   like mass accretion rate, Bernoulli's constant etc \cite{e}-\cite{j}. So far such works are done for non-self gravitating models of accretion, here we study the changes if the effect of the gravity of the accreting medium is considered. On the other hand, linear perturbations are introduced in such system to analyse the stability of the existing steady state solutions \cite{f}\cite{h}, we study the stability of such accretion models, and we find that there are some changes due to the effect of inclusion of the gravity of the medium. \\\\\\\\\\
\section{Spherically symmetric self-gravitating accretion model}
Two sources of gravity (Newtonian) are present, i.e. the gravity due to the massive accretor (the massive obeject, like star) and due to the medium itself. Fluid is falling steadily onto the accretor; due to such freefall, the flow is inviscid and irrotational and we assume  the fow to be spherically symmetric too. 
Therefore, the constant of motion arising from the Euler equation, i.e. Bernoulli's constant is given by \cite{d}\footnote{In this paper, the basic fluid equations are taken from this cited book.}\begin{equation}
\frac{v^{2}}{2}+\int \frac{dp}{\rho}+\Phi(r)={\rm constant},
\end{equation}
where fluid velocity is $v$, pressure is $p$ and $\Phi(r),~r$ being radial coordinate, is the gravitational potential due to the accretor and the infalling fluid itself. $\Phi(r)$ satisfying Poisson's equation is given below,
\begin{equation}
\Phi(r)=-\frac{GM}{r}-\frac{4\pi G}{r}\int_{R_\star}^{r}\rho r^{2}dr-4\pi G\int_{r}^{\infty}\rho(r)rdr,
\end{equation}
where $R_\star$ is the radius of the accretor.
The flow is barotropic. 
So our baratropic equation is
\begin{equation}
p=F(\rho).
\end{equation}
Thermodynamic sound speed can be defined as follows,
\begin{equation}
c_{s}^{2}=\left(\frac{\partial p}{\partial\rho}\right)=\frac{dF}{d\rho}.
\end{equation}
Similarly, for such steady flow, the conserved quantity arising from the continuity equation is mass accretion rate, given by
\begin{equation}
f=4\pi r^2\rho v.
\end{equation}
The steady state solution of this model is considered details in the articles \cite{b}\cite{c}.
\section{Linear perturbation in the system}
From now on, we denote the steady state velocity, density, pressure in the medium as $v_0(r),~\rho_0(r),~p_0(r)$ respectively.\\
Introducing linear perturbation in the flow, as
\begin{eqnarray}
& v(r,t)=v_0(r)+v(r,t)',\\
& \rho(r,t)=\rho_0(r)+\rho(r,t)'.
\end{eqnarray}
As pressure, $p$ is a function of density only, $p$ is also perturbed in linear order. 
Writing the full continuity and Euler equation as 
\begin{equation}
\frac{\partial \rho}{\partial t}+\frac{1}{r^2}\frac{\partial }{\partial r}(\rho vr^2)=0,
\end{equation}
\begin{equation}
\frac{\partial v}{\partial t}+v\frac{\partial v}{\partial r}=-\frac{1}{\rho}\frac{\partial p}{\partial r}-\frac{GM}{r^2}-\frac{4\pi G}{r^2}\int_{R_\star}^{r}\rho r^{2}dr.
\end{equation}
Defining a quantity, proportional to the mass accretion rate as 
\begin{equation}
\mathscr{F}=\frac{f}{4\pi}=\rho vr^2.
\end{equation}
In the steady state, $\mathscr{F}$ is a conserved quantity, $\mathscr{F}_0=\rho_0v_0r^2$, under linar perturbation,
\begin{equation}
\mathscr{F}=\mathscr{F}_0+r^2(\rho_0 v'+v_0\rho')=\mathscr{F}_0(r)+\mathscr{F}'(r,t).
\end{equation}
Therefore, in the continuity equation, equating the terms in the first order of smallness, we have
\begin{equation}
\frac{\partial \rho'}{\partial t}=-\frac{1}{r^2}\frac{\partial \mathscr{F}'}{\partial r}.
\end{equation}
Unlike the non self-gravitating case, after linearizing the Euler equation, we have terms from the gravitational field because the last term in equation (9) of the gravitational force, i.e. the gravitational force due to the medium, involves density. 
Now using the Euler equation in linear order and the above equation, we have a wave equation for $\mathscr{F}'$, given below
\begin{equation}
\partial_{\mu}(f^{\mu\nu}(r)\partial_{\nu})\mathscr{F}'(r,t)+\frac{4\pi G}{r^2}\mathscr{F}'(r,t)=0,
\end{equation}
where
\begin{equation}
f^{\mu\nu}(r)=\frac{v_0}{\mathscr{F}_0}\begin{bmatrix}
-1  &~~~ -v_{0} \\
-v_{0}&~~~ c_{s0}^{2}-v_{0}^{2}
\end{bmatrix}.
\end{equation}
$f^{\mu\nu}(r)$ can be related to the acoustic metric \cite{e}, \cite{f}, \cite{h} as emergent phenomena in the system. The timelike killing vector (as the acoustic metric is time independent), becomes spacelike as $v_0>c_{s0}$, i.e. after crossing the critical point or sonic point (in the case of transonic accretion solution) from subsonic region to supersonic region. Therefore, the radius at which $v_0=c_{s0}$ for transonic accretion, can be identified as 'dumbhole' horizon \cite{l}. The paper \cite{c} shows that the sonic point is shifted towards the accretor due to the inclusion of the gravity of the accreting medium, therefore the dumbhole horizon has smaller radius than that of in non-self gravitating transonic Bondi accretion solution. The second term in the wave equation is arising due to the inclusion of self-gravity; therefore, it can be considered as an interaction term between the perturbation in the medium with the gravity of the medium. The inverse square nature in the second term is due to the Newtonian inverse square gravity. The presence of this new term modifies the dispersion relation, using Eikanol approximation \cite{l} \cite{m}, we find
\begin{equation}
\omega=-v_0k\pm\sqrt{c_{s0}^2k^2-4\pi G\rho_0}.
\end{equation}
Therefore, for the stability of such high momentum wave propagation,
\begin{equation}
k\geq\frac{\sqrt{4\pi G\rho_0}}{c_{s0}}~\forall ~r~ {\rm in~ the~ flow}.
\end{equation}
 This criteria is very similar to the Jeans instability criteria \cite{d}, except for the fact that $\rho_0$ is a function of $r$. The stability criteria is also compatible with the Eikanol approximation of high momentum wave. To find the minimum possible $k$, $k_{min}$ for which the solution is stable, one has to investigate the function $\frac{\sqrt{4\pi G\rho_0}}{c_{s0}}$. For isothermal flow, $c_{s0}$ is constant and from the background solution of the accretion flow \cite{c}, $\rho_0$ is maximum at the surface of the star, $\rho_{0*}$. Therefore, for isothermal flow, for stability at all radii in the flow,
\begin{equation}
k_{min}=\frac{\sqrt{4\pi G\rho_{0*}}}{c_{s0}}.
\end{equation}
The corresponding value of maximum possible value of $\lambda$, $\lambda_{max}$ for stability,
\begin{equation}
\lambda_{max}=\sqrt{\frac{\pi c_{s0}^2}{G\rho_{0*}}}.
\end{equation}
For adiabatic flow, $p=F(\rho)=k\rho^{\gamma}$, where $k$ is a constant related to the specific entropy \cite{n}, $\gamma$ is the specific heat ratio; $\frac{4}{3}<\gamma<\frac{5}{3}$. Therefore, using equation (4), we get, the maximu possible value of $\lambda$ for all $r$,
\begin{equation}
\lambda_{max}=\sqrt{\frac{\pi\gamma k}{G\rho_{0*}^{2-\gamma}}}.
\end{equation}
 For instability, $\lambda_{max}$ has to be small enough so that the Eikonal approximation is also true.
Therefore, in both the cases instability arises from the medium near the surface of the accretor for waves  having wavelength above certain limit.\\
The expression of the group velocity from the dispersion relation,
\begin{equation}
v_g=\frac{\partial\omega}{\partial k}=-v_0\pm\frac{c_{s0}^2k}{\sqrt{c_{s0}^2k^2-4\pi G\rho_0}}.
\end{equation}
Therefore, relative to the moving medium, the group velocity of such wave is
\begin{equation}
v_g|_{medium}=\frac{c_{s0}^2k}{\sqrt{c_{s0}^2k^2-4\pi G\rho_0}}.
\end{equation}
Therefore, the dispersion relation is superluminal \cite{l}, i.e. the propagation speed of linear perturbation is greater than the thermodynamic speed of sound. For relatively high momentum wave, $v'_g$ is closer to the thermodynamic sound speed. Shorter wavelength perturbation has less speed. Therefore, the acoustic horizon is not absolute in the model, rather it depends on the wavelength of perturbation. The frequency dependence of the position of the acoustic horizon is also there in the quantum model of Bose-Einstein Condensate \cite{l}. Therefore, this is a classical analogue model of gravity where such thing is observed.  
\section{Linear Stability Analysis of stationary solution}
We consider a trial solution of the form \cite{o}, \cite{f}, $\mathscr{F}'(r,t)=\mathscr{F}_\omega (r) e^{i\omega t}$. We have from the equation (13),
\begin{eqnarray}
& \omega^2\mathscr{F}_{\omega}f^{tt}-i\omega [\partial_r(\mathscr{F}_\omega (r) f^{rt})+f^{tr}\partial_r \mathscr{F}_\omega (r)]\nonumber\\ 
& -\partial_r f^{rr}\partial_r\mathscr{F}_\omega (r)-\frac{4\pi G}{r^2}\mathscr{F}_\omega (r)=0.
\end{eqnarray} 
\subsection{Standing Wave}
A standing wave has vanishing amplitude of perturbation at two different radii, $r_1$ and $r_2~(r_1<r_2)$, i.e. $\mathscr{F}_\omega (r_1)=\mathscr{F}_\omega (r_2)=0$. Therefore, we have by integration
\begin{equation}
A\omega^2+C=0,
\end{equation}
where 
\begin{eqnarray}
& A=\bigintsss_{r_1}^{r_2}\mathscr{F}_{\omega}^2(r) f^{tt}(r)dr, \\
& C=\bigintss_{r_1}^{r_2}(\partial_r\mathscr{F}_\omega(r))^2f^{rr}(r) dr-4\pi G\bigintsss_{r_1}^{r_2}\frac{\mathscr{F}_\omega(r)^2}{r^2}dr.\nonumber\\
\end{eqnarray}
The expression of frequency is given by
\begin{equation}
\omega^2=-\frac{C}{A}.
\end{equation}
The above equation gives the frequency of such standing wave solution, i.e. the frequency depends on $r_1,~r_2$ and the standing wave profile. Hence
\begin{eqnarray}
& \omega^2=\frac{1}{\bigintsss_{r_1}^{r_2}\frac{\mathscr{F}_{\omega}^2(r)}{\rho_0 r^2}dr}\left[\bigintss_{r_1}^{r_2}(\partial_r\mathscr{F}_\omega(r))^2\frac{(c_{s0}^2-v_0^2)}{\rho_0r^2} dr\right]\nonumber\\
& -\frac{4\pi G}{\bigintsss_{r_1}^{r_2}\frac{\mathscr{F}_{\omega}^2(r)}{\rho_0 r^2}dr}\left[\bigintsss_{r_1}^{r_2}\frac{\mathscr{F}_\omega(r)^2}{r^2}dr\right].
\end{eqnarray}
The above expression implies that, if both the nodes (at $r_1$ and $r_2$) of such wave lie within the supersonic region of flow, the stationary flow is unstable under such perturbation which is also seen in the non self-gravitating case \cite{o}\cite{f}. In the supersonic region, it is not possible to spatially constraint the perturbation such that at two different radii, it vanishes at all time. In the subsonic region stability depends on the two competeting terms of the above equation.\\
 For stability,
\begin{equation}
\left[\bigintss_{r_1}^{r_2}(\partial_r\mathscr{F}_\omega(r))^2\frac{(c_{s0}^2-v_0^2)}{\rho_0r^2} dr\right]\geq4\pi G\left[\bigintsss_{r_1}^{r_2}\frac{\mathscr{F}_\omega(r)^2}{r^2}dr\right].
\end{equation}
The magnitude of $\mathscr{F}_{\omega}(r)$ has to be such that the assumption about the linearity of perturbation is valid. As $r$ increases, $\partial_r(\mathscr{F}_\omega(r))$ decreases due to stretching because we assume that between two radii, there is not a single point where, $\mathscr{F}_\omega(r)$ vanishes. Therefore, assuming that the two nodes are separated by a long distance, in the bulk region (region around the middle of the two nodes), $\partial_r(\mathscr{F}_\omega(r))$ is practically zero, making the integrand in the left hand side to zero. Assuming the integrals are finite at all points, if the separation between the two nodes increases, the left hand side of the above equation does not increase much as compared to the right hand side of the above equation. Therefore, after a certain limit (depending on $\mathscr{F}_\omega(r)$) of distance between the two nodes, stability would not be maintained. Hence, the stationary solution becomes unstable under such standing wave having wavelength greater than a certain length. Therefore, due to inclusion of self-gravity in the system, it is not possible to maintain standing wave with vanishing amplitude at large separation.
\subsection{Radially Travelling Wave}
The background stationary solution is smooth and continuous at all radii. The background quantities appearing in the linear wave equation, equation (22) are smoothly varying, hence we make use of WKB method \cite{p} to find solution for $\mathscr{F}_\omega(r)$. We seek solution of the form,
\begin{equation}
\mathscr{F}_\omega (r)=A(r) e^{i\theta(r)},
\end{equation}
where $A(r)$, the amplitude, is a slowly varying function of $r$, i.e. it varies slowly compared to $\theta(r)$ \cite{p}. Therefore, the solution gives more accuracy for short wavelength, i.e. in the high frequency limit.\\
Plugging such solution in equation (22), the real part and the imaginary part gives respectively
\begin{eqnarray}
& \omega^2 A f^{tt}+2A\omega f^{rt} \theta'-f^{rr}(A''-A\theta'^2)\nonumber\\
&-A'\partial_r(f^{rr})-\frac{4\pi GA}{r^2}=0,
\end{eqnarray}
\begin{eqnarray}
& \omega A \partial_r(f^{rt})+2\omega f^{rt} A'+f^{rr}(2 A'\theta'+A\theta'')\nonumber\\
&+A\theta'\partial_r(f^{rr})=0,
\end{eqnarray}
where $A'=\frac{dA}{dr},~\theta'=\frac{d\theta}{dr},~A''=\frac{d^2A}{dr^2},~\theta''=\frac{d^2\theta}{dr^2}$.
From equation (31), we find
\begin{equation}
A^2(\omega f^{rt}+\theta'f^{rr})={\rm constant}.
\end{equation}
Neglecting the term $A'',~A'\partial_r(f^{rr})$ in the equation (30) ($\because$ $A(r)$ is slowly varying and the background quantities are slowly varying with $r$ in comparision to $\theta(r)$), we find
\begin{equation}
\theta'=\frac{-\omega f^{rt}\pm \omega\sqrt{(f^{rt})^2-f^{tt}f^{rr}+\frac{4\pi Gf^{rr}}{\omega^2r^2}}}{f^{rr}}.
\end{equation}
Therefore,
\begin{equation}
A(r)=\frac{C}{\left(\omega\sqrt{(f^{rt})^2-f^{tt}f^{rr}+\frac{4\pi Gf^{rr}}{\omega^2r^2}}\right)^{\frac{1}{2}}},
\end{equation}
where $C$ is a constant depending on the initial condition on $\mathscr{F}_\omega(r)$.
Putting the values, we get
\begin{equation}
\theta'_{\pm}=\frac{\omega v_0\pm\omega\sqrt{c_{s0}^2+\frac{4\pi G\rho_0(c_{s0}^2-v_0^2)}{\omega^2}}}{c_{s0}^2-v_0^2}
\end{equation}
and
\begin{equation}
A(r)=C\left(\frac{\rho_0r^2}{\omega\sqrt{c_{s0}^2+\frac{4\pi G\rho_0(c_{s0}^2-v_0^2)}{\omega^2}}}\right)^{\frac{1}{2}}.
\end{equation}
Therefore, the complete solution can be written as
\begin{eqnarray}
& \mathscr{F}'(r,t)=\left(\frac{\rho_0r^2}{\omega\sqrt{c_{s0}^2+\frac{4\pi G\rho_0(c_{s0}^2-v_0^2)}{\omega^2}}}\right)^{\frac{1}{2}}e^{i\omega t}\nonumber\\
&\times \left[C_+e^{i\bigint^r \theta'_+(r)dr}+C_-e^{i\bigint^r \theta'_-(r) dr}\right],
\end{eqnarray}
where $C_+$  is the amplitude corresponding to the travelling wave, propagating radially outward and $C_-$ is the amplitude corresponding to the travelling wave, propagating radially inward. One can easily see this by dropping the term due to the self-gravity in the above equations. Interestingly, the amplitude of such wave is same as the amplitude of the travelling wave for non-self gravitating case at $v_0=c_{s0}$.
The expression of $\theta'_{\pm}$ is same as $k$ (in terms of $\omega$ and the background stationary quantities) in the equation (15). Therefore, the solution is stable for certain limit in $\theta'_\pm$. This is because, we are assuming high frequency wave; by WKB method, we get the expression of the amplitude in addition.\\
Alternatively, one can solve the equation (22) in high frequency limit by taking a trial solution of the form, given below
\begin{equation}
\mathscr{F}_{\omega}(r)=exp\left[\sum_{n=-1}^{n=\infty}\frac{K_n(r)}{\omega^n}\right].
\end{equation}
This standard method of travelling wave analysis used in several literature \cite{o}, \cite{f}-\cite{h}, would also give same solution found above. This is basically another way of implementing WKB method.

\section{Summary and Conclusions}
In this paper, we investigate the time dependent problem in perturbative approach in the non-relativistic spherically symmetric accretion model in Newtonian gravity. We find that the linear perturbation in mass accretion rate satisfies a wave equation with an interaction term with gravity due to the inclusion of gravity of the infalling medium. From the wave equation, we find the acoustic metric and we find that the interaction term with gravity in the wave equation, modifies the dispersion relation. Unlike the non-self gravitating case, the dispersion relation is superluminal, and as a result of this the acoustic dumbhole horizon is not absolute in our case, rather it's frequency dependent. This is a classical analogue model of gravity where such thing is observed. The inclusion of the gravity of the medium, gives rise to instabilities which are absent if the gravity of the medium is switched off.
 \section{Acknowledgements}
 The author is thankful to Prof. Sankhasubhra Nag of Sarojini Naidu College for Women, Kolkata, India for useful discussions and suggestions.

\end{document}